# Preservation Constraints on aDNA Information Generation and the HSF Posterior Sourcing Framework: A First-Principles Critique of Conventional Methods

Wan-Qian Zhao[1*], Shu-Jie Zhang[1], Zhan-Yong Guo[2], Mei-Jun Li[3]

1. School of Life Sciences, Zhengzhou University, Zhengzhou, China
2. College of Agronomy, Henan Agricultural University, Zhengzhou, China
3. National Key Laboratory of Petroleum Resources and Engineering, College of Geosciences, China University of Petroleum (Beijing), Beijing, China

[*]Corresponding author and email: wqzhao@zzu.edu.cn

**Abstract**

Fossil DNA preservation varies with depositional environments and diagenesis, producing fragments of heterogeneous origins and degradation states. We use first-principles biomolecular analysis to classify fossil molecular environments into four system types, distinguished by three indicator sets: origin (H/h: host/heterologous), deamination status (D/d), and similarity ratio (S/s). Conventional aDNA pipelines assume a binary mix of endogenous host DNA and modern contaminants, overlooking multisource complexity from multiple species and time-averaged deposits. This leads to bias: authentic signals suppressed during enrichment, alignment, or damage filtering, and exogenous/ancient admixed fragments misassigned as endogenous, particularly in open systems. We introduce the HSF (Host/Species-specific Fragment) posterior traceability framework to address this. It treats fragments as primary units, maximizes source diversity, detects isolated sequences, defers lineage assignment to preserve uncertainty, and applies phylogenetic consistency to discriminate origins. Combined with preservation characterization (e.g., 3D imaging and volumetric openness assessment), it improves authenticity evaluation and reduces misassignment in mixed-signal samples. Case studies identify novel fossil DNA patterns (CRSRR and SRRA) and demonstrate superior performance compared with conventional methods. The HSF framework enhances aDNA reliability, extends molecular archaeology to challenging contexts, and aids genome evolution and lineage reconstruction.

**Introduction: Frameworks for Authenticity Assessment in aDNA**

The central challenge in aDNA research stems from the fact that fossil formation and preservation span geological timescales. During this period, environmental variables (including temperature, pH, moisture content, mineral–organic interface reactions, and microbial activity) interact in heterogeneous and nonlinear ways [1–3]. These complex, site-specific diagenetic processes cannot be fully replicated by generalized degradation models derived from controlled laboratory experiments using purified DNA or from patterns observed in a limited set of exceptionally preserved samples [4–6].

Conventional aDNA methodologies implicitly assume that biomolecules recovered from fossils comprise a simple binary mixture: endogenous host residues plus modern contamination. Contamination mitigation therefore relies on physical cleaning, probe-directed enrichment, chemical damage markers (primarily cytosine deamination), and fragment-length filtering. In practice, the conventional workflow consists of three main steps. First, probe-directed enrichment selectively captures fragments complementary to the designed probes while excluding non-complementary sequences. This is followed by target-guided alignment using tools such as BLAST, MAFFT, or BWA, with reference databases restricted to the fossil taxon, the target species, and closely related genera. Finally, the results are filtered according to "classical aDNA features", primarily relying on terminal deamination patterns and fragment-length thresholds (typically ≤60 bp) [7, 8].

This paradigm, however, overlooks a critical aspect of fossil taphonomy: exogenous biomolecules can infiltrate fossil interiors during diagenesis through pore networks, groundwater flow, or hydrocarbon fluids, leading to long-term mixing with endogenous molecules [9-13]. Phylogenetically biased probes applied early in the workflows introduce systematic bias, which is further amplified by subsequent filtering steps. Fragments from environmental, parasitic, or symbiotic sources that happen to share sequence similarity or degradation signatures with the target may be preferentially retained, while authentic host-derived fragments lacking these diagnostic features are disproportionately discarded. Taken together, these steps amplify structural limitations in conventional approaches, particularly in non-closed preservation systems where multisource molecular assemblages are common.

From a first-principles perspective, we introduce the HSF selection framework to systematically address the wide range of preservation conditions and corresponding biomolecular decay patterns observed in fossils. The framework treats individual sequence fragments as the fundamental unit of analysis to reduce methodological bias introduced by early experimental priors. It integrates degradation kinetics with unbiased sequence-based evidence to support more robust authenticity assessment and enables direct, systematic comparison between results from conventional pipelines and HSF-based processing. Potential broader applications include large-scale HSF screening to inform genome evolution inferences and taxon-specific genome reconstruction. Although developed primarily for aDNA, the approach delineates general principles for information recovery under severe preservation constraints and provides a conceptual foundation for studying complex molecular depositional environments.

## 1. From Time Capsules to Depositional Systems

### A System-Level Framework for Molecular Preservation

To address the complexity of the "fossil state–biomolecular decay" relationship, we propose a classification of four representative preservation systems (non-exhaustive) based on commonly observed patterns of molecular preservation in fossils, delineated along a physical gradient of molecular openness and exchange potential:

> **Closed-preserved System:** biomolecules are rapidly isolated from water after burial—typically by encapsulation in minerals or lipids, or by permanent freezing. As a result, chemical degradation is negligible and endogenous biomolecules remain largely intact. Although a perfectly closed-preserved fossil system has not been empirically confirmed, localized or transient examples may exist (e.g., asphalt seeps, certain siliceous concretions, and salt-cave deposits).
>
> **Closed-degrading system:** This system is characterized by extremely low porosity and near-complete isolation from the external environment. Residual water phases induce slow, time-dependent chemical degradation. Many high-profile ancient human genomic datasets reported in the literature likely belong to this category, although their classification requires further validation within a system-level framework [14-16].
>
> **Open-preserved system:** Here, porous media are subject to repeated cycles of wetting and drying as well as groundwater flow, resulting in continuous water-mediated material exchange [9-11]. This leads to persistent introduction and long-term accumulation of exogenous molecules, producing complex molecular mixtures following external input. Certain fossils from Jehol Biota provide representative examples of this system [17].
>
> **Invasion-replacement system:** In this scenario, infiltration by microorganisms, small animals, or plant roots results in the predominance of exogenous biomolecules that coexist with endogenous components, ultimately altering the overall molecular composition [16, 17]. Instances of extensive parasitic colonization within large fossil bones illustrate this type.

Additionally, a fossil may belong to one preservation system during a particular geological period but can subsequently transition to another system through geological processes (termed "**system transition**"). Different regions of the same fossil may exhibit distinct preservation types simultaneously due to variations in composition, structure, or local burial microenvironment (termed "**system coexistence**").

These categories represent common modes of molecular exchange rather than mutually exclusive states. Therefore, differences in biomolecular preservation should be interpreted on the basis of the physicochemical properties of the sampling location and the potential for inter-system transitions during diagenesis, rather than chronological age alone. This classification is defined by boundary conditions governing material exchange rather than by specific environmental types, thereby ensuring greater descriptive completeness at the system level.

---

## 2. Fossils as Mixed Molecular Reservoirs

### Framework of a three binary-pair framework

In non-closed systems, the assumption that fossil DNA originates solely from the host, with any other DNA being modern contamination, does not hold. Instead, a fossil should be regarded as a molecular depositional reservoir that evolves continuously over time and contains biomolecules derived from multiple periods and multiple species. To systematically classify fossil biomolecules, we introduce a three binary-pair framework, which categorizes fragments according to three independent attributes:

**Host (H)** vs. **non-host (h)** origin.

**Deaminated (D)** vs. **non-deaminated (d)** status.

**Similarity** (S) vs. **dissimilarity (s)** to the target genome.

Here, the "target genome" refers to the reference sequence used during probe enrichment and sequence assembly. Early ancient human studies typically used the modern human reference genome (e.g., rCRS), whereas more recent work often employs assembled Denisovan or Neanderthal genomes.

This system allows fragments to be represented using three-letter combinations (e.g., HDS, hds), facilitating quantitative comparison of preservation and origin properties. Consequently, eight molecular categories arise: HDS, HDs, HdS、Hds, hDS, hDs, hdS, and hds.

**The minimal complete state space of fossil biomolecules**

This classification is not empirical but is derived from the current information framework as an operationally minimal and complete state space for lineage attribution.

The relative proportions of these molecular categories are determined by three physical variables:

**Exogenous input load (H/h)**

This reflects the balance between externally introduced molecules (via water or other transport pathways) and endogenous host biomolecules, including contributions from symbiotic or parasitic organisms.

**Similarity expression ratio (S/s)**

This represents the proportion of sequences that are similar versus dissimilar to the target genome, reflecting the degree of external sequence input relative to host-derived signals.

**Water-phase contact (D/d)**

This indicates whether molecular fragments have experienced aqueous exposure within the sample. Such exposure directly affects water-mediated chemical degradation rates, including deamination.

These three variables jointly determine both the state and evolutionary trajectory of preservation systems.

**System–molecule correspondence**

In summary, at the macroscopic scale, preservation can be classified into four system types, while at the microscopic level this corresponds to eight molecular state categories. Their distributions are not equiprobable; instead, their relative proportions constitute diagnostic signals of preservation system states.

Accordingly, deamination alone should not be treated as the sole discriminative axis; physical system variables regulate the steady-state proportions of all eight molecular categories; these proportional relationships influence lineage attribution outcomes. Thus, the compositional profile of the eight molecular classes simultaneously reflects preservation system status and constrains subsequent origin inference. A detailed discussion of the relationships between preservation systems and the eight molecular classes is provided in the first section of Materials and Methods.

---

## 3. Logical Limitations of Conventional Methodologies

Conventional methods often ignore that multiple systems can coexist within fossils and evolve over time, particularly in open systems or those subject to external intrusions. When eight molecular categories are mixed simultaneously, the number of distinguishable molecular states is mathematically compressed, which can introduce structural bias into analyses. Figure 1 shows the screening process for these eight molecular categories.

### 3.1 Systematically Discarded Environmental and Ecological Signals

Probe-directed enrichment (kinship-based/target capture) preferentially recovers fragments complementary to the designed probes, typically from a specific genus, species, or reference genome; thus, systematically excludes non-complementary fragments. Consequently, informative signals from other taxa (environmental/exogenous sources) are often discarded early, creating systematic information bias.

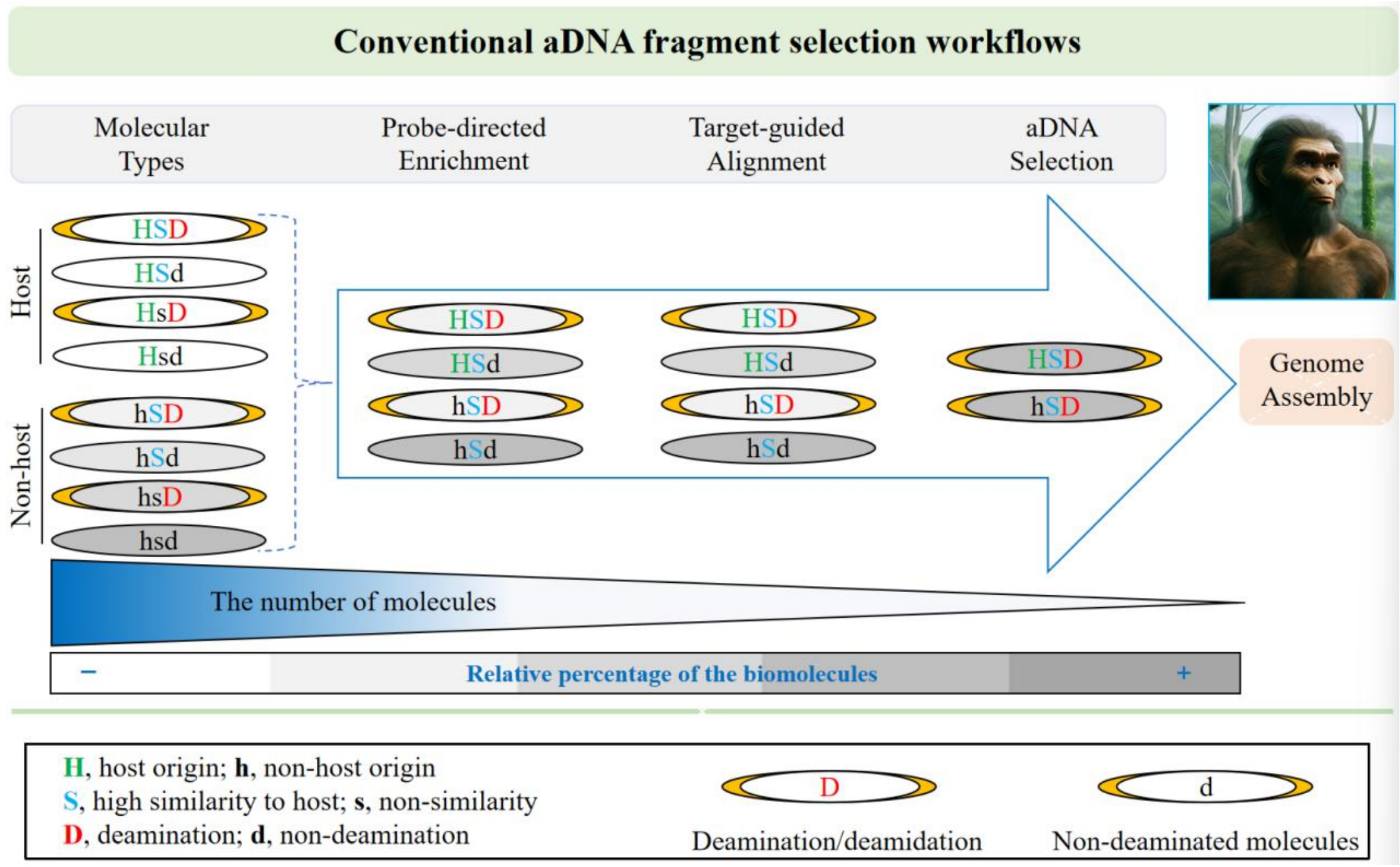

Figure 1. Effects of three critical steps in the conventional workflows on eight categories of ancient hominin molecules. The pipeline is: probe-directed enrichment → target-guided alignment → deamination filtering. Probe enrichment led to loss of Hsd/HsD/hsD; sequence alignment biased fragment selection toward fossil taxa and misidentified hSD; deamination filtering missed HSd, causing downstream SNP-calling and assembly errors. Causal chain: probe complementary enrichment → sequence alignment parameter shift → deamination-filter omission → systematic misselection/omission of large molecules.

### 3.2 Loss of Host Biomolecules

Biomolecular degradation involves both detectable and undetectable forms of damage. Undetectable damage includes oxidative modification, crosslinking, Maillard reactions, and degradation into ultrashort fragments. Detectable damage primarily involves two pathways under nuclease-free, neutral pH, ambient conditions:

**Depurination** followed by **β-elimination**, leading to strand breaks

**Cytosine deamination** (C→T transitions)

The depurination → β-elimination pathway occurs at approximately 2.4 times the rate of deamination [18-20]. Furthermore, strand breaks generated by depurination can result in the loss of terminal nucleotides, including those bearing deamination signatures. At present, no mathematical model adequately describes the temporal evolution of deamination in DNA fragments.

Thus, fragments displaying deamination patterns represent a survivorship-biased subset of the original host DNA population, whereas a substantial fraction of authentic host fragments lacking detectable deamination are systematically excluded.

Based on *in vitro* parameters, in the host DNA filtering system employed by the Fu–Pääbo group

(2025), non-deaminated host fragments may outnumber fragments exhibiting six terminal C→T substitutions by approximately 14.4-fold [7]. Although this estimate derives from *in vitro* parameters and has limited precision, the conclusion that large numbers of host fragments are discarded remains robust.

It should also be noted that when biomolecules are isolated from water, both depurination and deamination rates decrease dramatically or effectively cease.

Throughout this work, the term "deamination" carries three distinct meanings depending on context: A molecular state variable; a system-level preservation indicator; an analytical filtering parameter. These meanings must be distinguished carefully.

### 3.3 Fragment Misassignment: Overlooked Critical Factors

Conventional workflows prioritize fragments with high similarity to the target sequence or its degradation signatures (terminal deamination, short length), while discarding signals lacking these features. This selection logic is vulnerable to systematic misassignment for the following reasons:

**Improper use of chemical indicators:** deamination and strand breakage depend on local diagenetic conditions (pH, temperature, aqueous exposure, salinity, mineral interfaces). Damage varies significantly within the same fossil, rendering uniform thresholds unreliable for inferring antiquity or separating endogenous from exogenous material.

**Failure to account for multisystem coexistence:** a single fossil can contain molecules from multiple temporal phases and taxa. Ancient exogenous DNA may acquire degradation signatures indistinguishable from host fragments, leading to unresolved overlapping H/h signals that conventional pipelines cannot disentangle.

**Systematic bias from target-guided sequence alignment:** when the search scope is limited to close relatives, non-host-specific sequences (especially conserved interspecific regions of 30~60 bp) are particularly prone to being misclassified as host-derived sequences.

**Neglect of historical contamination sources:** in sites with prolonged human activity or complex taphonomy, diagenetic processes can continuously introduce and mix biomolecules from diverse ancient origins (including environmental *Homo* biomolecules) unless system closure is demonstrated. Conventional filtering cannot reliably exclude fragments that share highly similar sequences or chemical signatures but originate from different periods, individuals, or taxa.

**These limitations create a structural methodological barrier**: without independent confirmation of preservation-system closure, prevailing approaches cannot robustly distinguish endogenous host signals from ancient exogenous or multisource inputs.

### 3.4 Ancient Human Genomes as Intrinsically Uncertain Outcomes

Reliable genome assembly from ancient human material requires the simultaneous satisfaction of the following conditions: high endogenous content, low levels of removable contamination (quantifiable, for example, via h-HSF metrics; see Materials and Methods), sufficient fragment length distribution and precise mappability, appropriate selection of reference sequence, demonstration that SNPs are independent of postmortem chemical damage, effective exclusion of fragmentation uncertainties (including M-type, P-type, and AT-type, see Materials and Methods). If any of these conditions is not met, genome assembly should be deferred or avoided to prevent systematic fragment misselection or omission.

In studies of ancient humans, the vast majority of skeletal remains are non *in situ*, usually incomplete skeletons or bone fragments, and are therefore typically preserved in open depositional systems. No matter how parameters are optimized in conventional workflows (e.g., by expanding reference datasets of sequence alignment, increasing probe specificity, or adjusting hybridization thresholds), it remains extremely difficult to reliably distinguish, on the basis of sequence similarity or chemical damage signatures (e.g., deamination patterns), fragments originating from different temporal phases, different individuals (including the presumed host, other archaic hominins, modern humans, or non-hominin taxa) that exhibit highly comparable characteristics to the target sequence.

This constitutes a fundamental logical barrier inherent to conventional methodologies. The only robust way to resolve this impasse is to perform independent assessment of preservation-system closure prior to downstream analysis. In the absence of evidence confirming that the sample represents a near-closed system, the resulting data carry intrinsic uncertainty, and any associated biological or evolutionary inferences, including gene flow between archaic and modern humans, admixture timing and direction, and proposed functional links between archaic alleles and modern phenotypes or disease risk, must be regarded as provisional and subjected to systematic reevaluation.

### 3.5 The True Value of Chemical Damage Markers

Deamination is primarily controlled by liquid water availability, salinity, and temperature. It therefore reflects molecular contact with aqueous phases rather than molecular age itself. Accordingly, deamination is better interpreted as an indicator of preservation environment and exposure history (such as wetting frequency, pore heterogeneity, and geochemical conditions), rather than as a direct proxy for molecular antiquity. aDNA may exhibit deamination or remain unaffected depending on environmental conditions. Thus, deamination should be regarded as an environmental response variable, useful for reconstructing exposure histories and contextualizing damage patterns across fossil matrices.

### 3.6 Data Traceability Limitations

Conventional workflows often lack adequate data traceability. Published genome assemblies frequently omit detailed contig-level information, including fragment identifiers and damage patterns, which should ideally be archived as accessible supplementary materials.

Because traditional pipelines do not treat fragment-level provenance tracking as essential, manual verification becomes extremely difficult when dealing with large raw datasets.

### 3.7 Valid Scope of Conventional Methods

It is important to emphasize that this study does not deny the effectiveness of deamination-oriented approaches in highly closed systems characterized by minimal material exchange and dominance of a single host source. However, these methods implicitly generalize the assumption of a reducible closed system to arbitrary systems. When applied to open or multi-system samples, systematic biases are likely to arise.

Until preservation system classification and molecular origin are clearly established, conclusions based solely on conventional methodologies should be regarded as containing inherent uncertainty.

---

## 4. HSF Selection: A New Analytical Framework for Ancient DNA Fragments

To avoid a priori assumptions and resolve fragment provenance in the presence of overlapping H/h signals, we propose a streamlined signal-separation framework consisting of the following core principles and procedures.

### 4.1 General Analytical Strategy

To avoid introducing target-specific bias during data generation, library preparation is performed without probe-directed enrichment or other prior assumptions about the expected host taxon. In multisystem depositional contexts where non-host signals may dominate or interfere, the framework prioritizes the explicit identification and classification of HSFs.

Two principal categories are distinguished:

H-HSF — fragments reliably attributable to the presumed host individual or taxon;

h-HSF — fragments attributable to exogenous sources, other taxa, or post-depositional intrusion (treated as potential contamination or mixed inputs).

These HSFs serve as the primary indicators for fragment provenance and traceability. Chemical damage patterns (e.g., deamination) are retained for conditional use when preservation-system context makes them informative, but are not applied as a universal authenticity filter.

### 4.2 Fragment Attribution Principles

Fragment origin is determined based on intrinsic sequence properties, including: Similarity to candidate host genomes, statistical significance of alignment, consistency with established phylogenetic relationships.

Specifically: HSD/HSd categories are used to extract candidate host signals; hSD/hSd categories are used to identify environmental (non-host) signals; the remaining molecular classes are used to characterize preservation system states and inform modeling; detailed operational procedures are provided in the Materials and Methods section.

---

## 5. Paradigm Differences Between Conventional Methods and HSF Selection

### 5.1 Conventional Approaches

From a logical standpoint, conventional methodologies implicitly assume a reducible closed–degrading system and adopt a species-prior framework. As mentioned before, from targeted probe enrichment to the subsequent selecting steps, resulting experimental outcomes do not fully align with the actual complexity of fossil molecular systems.

From a procedural perspective, conventional workflows rely on probe-directed enrichment, during which exogenous fragments may be inadvertently captured. When combined with high-cycle PCR amplification (typically ≥35 cycles) and target-guided alignment, both the experimental and computational stages become strongly biased toward the predefined target, thereby amplifying early errors. These steps collectively exacerbate structural blind spots in the traditional pipeline.

### 5.2 HSF Selection

From a conceptual standpoint, it treats fossil biomolecules as existing in a multi-system stochastic state space defined by H/h × D/d × S/s. The framework first evaluates statistical significance in phylogenetic distribution and subsequently imposes constraints based on phylogenetic consistency.

Methodologically, it follows a Bayesian-iterative logic: uncertainty is explicitly represented, iteratively updated, and progressively constrained throughout the analysis. Operationally, it employs standard library preparation (typically 5–20 PCR cycles) without target-specific enrichment or reference restrictions during sequencing and alignment. Fragment provenance is determined post hoc based on statistical significance and phylogenetic consistency. This approach preserves multispecies signals, enables concurrent detection of SNPs and complex recombination structures, and substantially expands the recoverable information space.

### 5.3 Handling Incomplete Reference Databases in the HSF Framework

When reference databases are incomplete (e.g., when certain taxa are underrepresented), alignment to the phylogenetically closest sequenced species should be treated as an explicit, predictable projection bias. In the HSF framework, this is managed by recording and reporting projection uncertainty, incorporating projection weights or conducting sensitivity analyses in downstream models as appropriate, and conservatively elevating assignments to the next stable phylogenetic rank (e.g., family) when genus-level data are absent.

Reference incompleteness thus becomes a controlled, transparent analytical parameter rather than a hidden bias amplified by early experimental design.

### 5.4 Two Distinct Conceptual Frameworks for Fossil Biomolecules

The divergence between the two conceptual frameworks arises from differing assumptions about the fundamental problem of how to represent molecular assemblages in fossils (Figure 2).

The HSF framework advocates characterization along at least three orthogonal dimensions (H/h, S/s, D/d). Within this approach, the S/s axis should adopt a broad taxonomic scope that encompasses all possible biological sources contributing to the molecular assemblage, rather than being restricted to phylogenetically close relatives. Relying solely on the D/d dimension to dichotomize molecular origins into “modern” versus “host-age-equivalent” categories is insufficient to capture the complex interplay between species provenance and temporal context. Additional perspectives, including preservation-system state, paleogeographic setting, climatic evolution, geological events, lineage continuity, and ecological dynamics, are therefore required to adequately reconstruct time–provenance relationships.

Fragments recovered by the HSF workflow may be deaminated or non-deaminated and are not subject to length restrictions (fragments >100 bp are routinely retained). These fragments carry relatively rich genetic information, and some can support reliable phylogenetic inference even without assembly into longer contigs.

In contrast, conventional approaches typically reduce the representation to two dimensions (S/s and D/d). In this framework, the S/s axis is often restricted to genus- or species-level phylogenetic alignment, while the D/d axis is coarsely partitioned into “modern-origin” and “host-age-corresponding” categories. Subsequent analytical steps further compress or discard time–provenance information, resulting in substantially reduced representational power and diminished capacity to distinguish correct assignments from erroneous conclusions. The “classical aDNA features” under the conventional perspective are short fragments (approximately 30~60 bp) exhibiting terminal deamination. Such fragments carry limited genetic information and are associated with a high risk of misassignment.

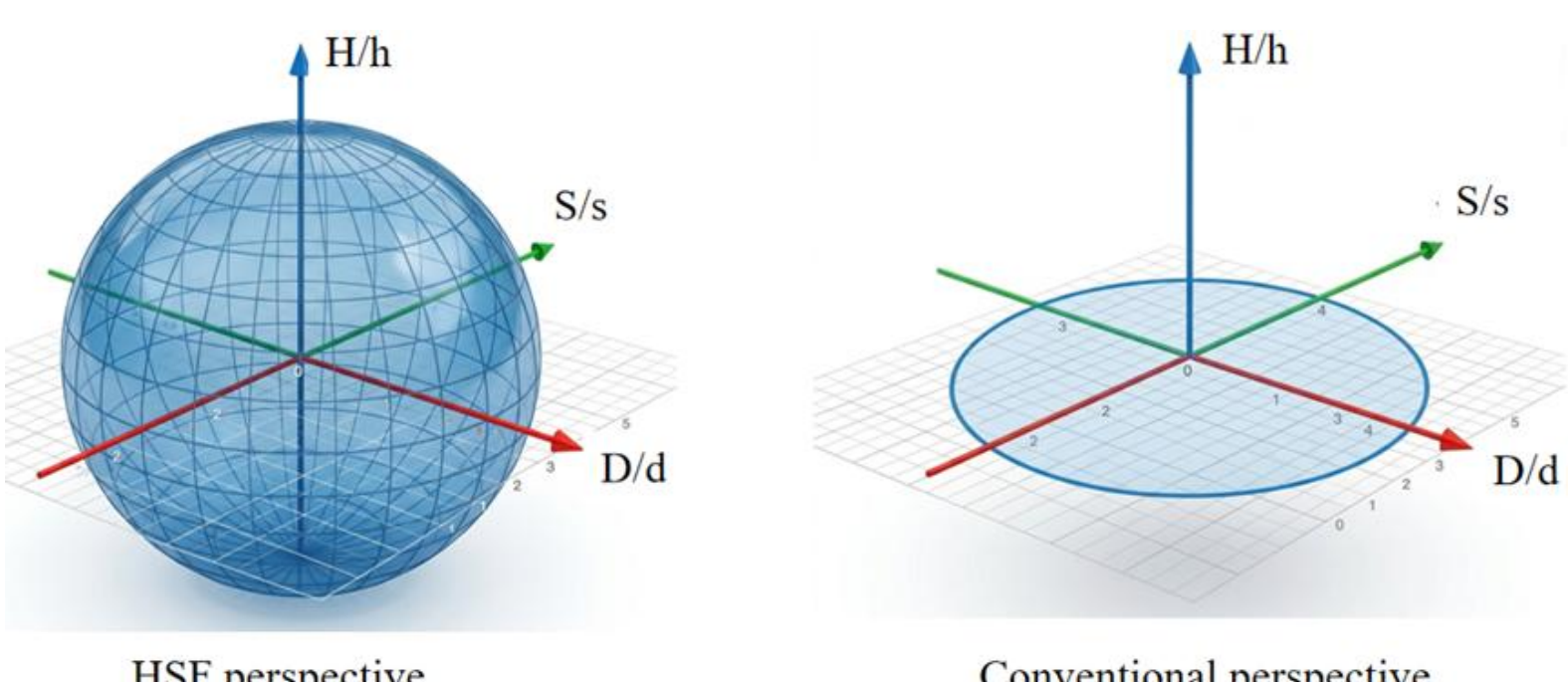

Figure 2. Two perspectives on fossil biomolecular assemblages. In the three-dimensional space of the HSF framework, the eight molecular categories can be fully distinguished and uniquely determined by the corresponding constraints. In contrast, within the two-dimensional space of the conventional framework, these eight categories cannot be completely resolved, nor does there exist a set of constraints capable of producing a one-to-one mapping. The H/h, S/s, and D/d axes are shown as orthogonal, but they are not necessarily required to be.

### 5.6 Validation of HSF Selection and Conventional Methods

HSF Selection employs a standard next-generation sequencing workflow without target-specific enrichment, whereas conventional methods rely on probe-directed enrichment followed by target-guided amplification strategies. The two approaches therefore differ systematically in enrichment bias, fragment representativeness, and patterns of error accumulation.

The core distinction between the methods does not lie in whether raw data should be forced into a single "universal" format compatible with both pipelines (since the data are generated under fundamentally different conceptual premises and their interpretation depends on post hoc fragment alignment and probabilistic inference). Rather, the critical difference concerns the prior assumption about the nature of the fossil: whether it is treated as a dynamic reservoir capable of continuously accumulating biomolecules from multiple temporal phases and biological sources. This conceptual prior fundamentally shapes the experimental design and analytical logic, rendering the two workflows products of mutually incompatible inferential frameworks.

Ideally, rigorous comparison would involve parallel application of both methods to identical specimens or reanalysis of previously generated non-enriched raw libraries using the HSF framework. However, large-scale systematic comparisons remain currently unfeasible due to the absence of mandatory cross-laboratory sample-sharing protocols and the scarcity of dedicated funding for validation studies.

---

## 6. Empirical Case Studies

### 6.1 Limitations of Deamination as an Authenticity Criterion

At present, conventional analytical workflows are primarily applied to fossils younger than approximately one million years and are typically limited to SNP detection at species or subspecies resolution. Such approaches provide relatively limited insight into broader paleoenvironmental and ecological information. Based on methodological inference, these workflows are most suitable for fossils preserved under relatively stable physicochemical conditions, particularly closed–degrading systems characterized by limited material exchange and intermittent water supply.

For fossil samples confirmed as closed systems — through three-dimensional X-ray imaging and internal volumetric ratio analysis (IVR)—no deaminated fragments have been observed. Examples include well-preserved fossil specimens containing primate and plant aDNA fragments recovered from sealed containers [17]. In such cases, biomolecules were rapidly isolated following deposition and remained protected from subsequent aqueous exposure. These observations demonstrate that deamination is a conditional and localized phenomenon rather than an intrinsic property of aDNA, consistent with established degradation kinetics.

### 6.2 Empirical Example 1

**Re-evaluation of Fragments Screened by Conventional Methods**

To assess the performance of conventional methods on real data, we selected published ancient hominin genome datasets and associated figures/tables. Reads/contigs annotated as *Homo* were subjected to independent re-analysis.

(1) Non-hominin Top Matches (error): Several fragments annotated as "*Homo*" were extracted from supplementary figures/tables or from assembled genomes in multiple published studies and re-aligned against the NCBI nr/nt database. Results showed that, for some fragments (e.g., row 2 in Table 1), the top three highest Ident (α) matches were microbial sequences rather than hominin. Other fragments exhibited M-type or P-type uncertainty (see Materials and Methods). These findings indicate that traditional screening pipelines may fail to effectively exclude exogenous or symbiotic sequences.

(2) Alignment Bias in Re-assessed Sequencing Data: Public sequencing datasets (e.g., ERR922054 and GWHAMOS01000001) were re-aligned using an unbiased BLAST online search against the NCBI nr/nt database. The fractions of fragments without certain hits were 4.0% and 3.6% for the two datasets; M-type uncertain fragments accounted for 76% and 49.9%, respectively, and P-type uncertain fragments were 6% and 17.9%. Related species included modern humans, human commensal bacteria, environmental bacteria (e.g., *Lycogala flavofuscum*), and environmental arthropods (e.g., *Oppiella nova*). The results indicate that conventional methods may have already introduced substantial distortion into their original data, a risk that has previously been overlooked.

(3) Concealment in the Assembly Process: In the progression from raw sequencing data to contigs and ultimately to final genome assembly, most studies fail to disclose the specific fragment origins used in the assembled genomes, including read accessions and the sequences themselves. Assembly parameters are often undocumented or inadequately described, resulting in a lack of transparency in critical selection steps.

In the case of the Harbin individual study, the key dataset Harbin_0.3 comprises nearly two million sequencing reads (requiring gigabyte-scale files for full representation), yet only 2047 unique sequences were ultimately selected for assembly. These could be clearly presented in a simple FASTA file of only a few kilobytes, but neither the read accessions nor the sequences of these 2047 fragments are disclosed in the publication [7]. This “one-in-a-thousand” low-probability selection combined with information opacity makes it extremely difficult for external researchers to independently verify or reproduce the critical steps of the analysis.

**Table 1. Uncertainties in identifications produced by conventional methods**

| Source | Sequence | bp | Scientific Name | Cover | Per. Ident | Ident (α) | E value | Accession | M/P |
|---|---|---|---|---|---|---|---|---|---|
| Suppl. Fig. 4 (a) ; Denisova Cave, a fragment of a phalanx [21] | TGCTTACAAGCAAGCACAACAATCAACCCTCAACTATCACACA<br>TCAACCGTAACCCCAAAGCCAACCCTCATCCACTAGAATATCA<br>ACAAACCTACCCATCCTTAACAGCACATAGCACATACAGTCAT<br>TTACCGTACATAGCACATTACAGTCAAATCCTCTCTCGCCCC<br>CACGGATGACCCCCCTCAGATAGGGATCCCTTGG | 205 | *Homo sapiens* | 100% | 93.63% | 93.63% | 1.00E-78 | EU565882.1 | P |
| | | | *Oppiella nova* | 100% | 90.69% | 90.69% | 4.00E-68 | OC916209.1 | |
| | | | *Lycogala flavofuscum* | 100% | 90.69% | 90.69% | 4.00E-68 | AY743240.1 | |
| | | | *Pan troglodytes* | 100% | 84.95% | 84.95% | 1.00E-48 | KX211943.1 | |
| Fig. S8.6: second read; Denisova Cave, a fragment of a toe bone [22] | GTTAGGGTTAGGGTTAGGGTTAGCTAACCCTAAACCCTAACC<br>CCTAACCC | 50 | *Abramis brama* | 100% | 96.00% | 96.00% | 3.00E-11 | OZ023079.1 | Error † |
| | | | *Arianta arbustorum* | 100% | 96.00% | 96.00% | 9.00E-11 | OZ317013.1 | |
| | | | *Boreogadus saida* | 100% | 96.00% | 96.00% | 3.00E-11 | OZ177913.1 | |
| | | | *Homo sapiens* | 100% | 94.12% | 94.12% | **2.00E-12** | M73018.1 | |
| C_AA108316.1; Harbin non-*in situ* buried dental calculus fragments [7] | ACTCACGGGAGCTCTCCATGCATTTGGTATTTTCGTCTGGGN<br>GGTGTGCA | 50 | *Homo sapiens* | 100% | 98% | 98% | 2.00E-14 | OL344174.1 | M |
| | | | *Oppiella nova* | 100% | 98% | 98% | 2.00E-14 | OC916209.1 | |
| | | | *Pan troglodytes* | *96%* | 98% | 94.00% | 2.00E-13 | AC189738.3 | |
| | | | *Gorilla gorilla* | 96% | 97.92% | 94.00% | 2.00E-13 | KM510341.1 | |
| PRJCA002765; Baishiya Karst Cave Pleistocene sediments [23] | CTTCTAACCACAGCACNTAAACATATCTCTGCCAAACCCNAA<br>AAACAAAGAACCCTAACACCAGCCTAGCCAGA (=layer 2 d) | 74 | *Homo sapiens* | 100% | 0.9459 | 94.59% | 5.00E-23 | KR698932.1 | P |
| | | | *Zea mays* | 100% | 91.89% | 91.89% | 9.00E-20 | EU949474.1 | |
| | | | *Oppiella nova* | **100%** | 91.89% | 91.89% | 9.00E-20 | OC916209.1 | |
| | | | *Semnopithecus entellus* | 91% | 91.04% | 82.85% | 7.00E-16 | EF053399.1 | |
| | CCAAACCATTTACCCAAATAAAGTATAGGCGATAGAAATTGT<br>AACCTGGCGCAATAGATATAGTACNGNAAGGGAAAG (=layer3<br>d) | 78 | *Homo sapiens* | 100% | 96.15% | 96.15% | 6.00E-27 | FJ383614.1 | M |
| | | | *Klebsiella pneumoniae* | 100% | 96.15% | 96.15% | 6.00E-27 | AP025034.1 | |
| | | | *Oppiella nova* | 100% | 96.15% | 96.15% | 6.00E-27 | OC916209.1 | |
| | | | *Pan paniscus* | 100% | 94.87% | 94.87% | 3.00E-25 | JF727228.2 | |
| | GTCAAGGTGTAGCCCATGAGGTGGCAAGAAAT**A**GGCTACATT<br>TTCTACCCCAGAAAACTACG (=layer7 d)<br>CGTAGTTTTCTGGGGTAGAAAATGTAGCC**T**ATTTCTTGCCAC<br>CTCATGGGCTACACCTTGAC-- Revised | 62 | The A or T base variants, emphasized in bold red underlines, can be challenging to distinguish whether these bases result from deamination or mutation. | | | | | | AT |

† Best match: non-hominins

More importantly, when misassignments occur at early stages (e.g., exogenous fragments erroneously aligned), subsequent contig- or scaffold-level alignments frequently converge toward the human reference genome, producing an apparently plausible archaic hominin profile that systematically conceals underlying biases and contamination effects.

Even in many published figures and processed assembly results (already curated datasets), non-human alignment signals remain detectable (see Table 1). This indicates that the problem stems not only from inherent methodological limitations but also from discretionary curation choices and selective disclosure practices. These factors jointly amplify the risk of error while simultaneously obscuring its presence through data and procedural opacity.

(4) Unknown Factors and Potential Mixed Signals: When methodological limitations are combined with multiple unknown influences (e.g., burial environment heterogeneity, symbiotic/parasitic sequence interference, sample source variability), resulting inferences may deviate from biological reality.

In the study of dental calculus from the Harbin cranium (>146,000 years old), two mtDNA sequences reconstructed from the same sample (GenBank accessions: C_AA108315.1 and C_AA108316.1) exhibit T/C heteroplasmy at rCRS position 14073 [7]. This suggests DNA from multiple individuals infiltrated the fossil, producing overlapping fragments from different time periods or donors during burial and diagenesis. Consequently, the assembled genome represents a composite mixture rather than the genetic material of a single individual. It underscores the risk of host sequence mixing with exogenous or multi-temporal DNA in open-system samples.

### 6.3 Empirical Example II

**Capturing Deterministic Signals Amid Uncertainty: Multispecies Fragments and Two Novel Molecular Structures in a Lycoptera Fossil**

As a case study from the Jehol Biota, 3D X-ray imaging combined with internal volumetric ratio (IVR) testing of a *Lycoptera* fossil revealed the coexistence of closed and open preservation systems within the same specimen. The sequencing data contain numerous fragments derived from plants, fish, and mammals (see PRJNA1309836). Many sequences project onto modern genomes with generally low Ident (α) values, indicating that they are unlikely to derive from extant species. When considered in the context of paleoenvironmental, ecological, biogeographic, and evolutionary factors, these fragments are more plausibly interpreted as ancient exogenous inputs (see Table 2).

**Table 2. HSF selection of DNA fragments from *Lycoptera* fossils**

| Source | Sequence | Bp | Origin | Scientific Name | Cover | Per. Ident | Ident (α) | E value | Accession | Deamination |
|---|---|---|---|---|---|---|---|---|---|---|
| Table S3 [17] | AGGGAGATGGGAATTATTCCTAAGGTTGAGCCGATGTAGAGTGTGGCGCCCTCGCTTGTATATGTGTAAAGGCTTGTACGATGAAGTTGTGTGCCCTCGCTTGCCTGTGCCTCGCTTGCTTGTACGATGAAGGTTGAGCCGTTGTGTACT | 150 | *Zea mays* : 5 fragments | *Zea mays* | 75% | 99.11% | 74.33% | **1.00E-47** | KT989678.1 | N/A |
| | | | | Non-matched | | | | | | |
| Table S1A [17] | AGAGACGGGGTTTCACCGTGTTAGCCAGGATGGTGTCGATCTCCTGACCTCGTGATCCACCTGCCTCAGCCTCCCAAAGTGCTGAGATGACAGGCGTGAGGCACCGCACCAGTCCTTTTTTTTTTTATTATTTGAGAATGAGTTTCACTA | 150 | Non-human primates: 50 fragments | *Pan troglodytes* | 95% | 93% | 88.36% | **8.00E-51** | CT005239.3 | N/A |
| | | | | *Homo sapiens* | 99% | 90% | 89.10% | 2.00E-47 | AL590379.7 | |
| | | | | *Gorilla gorilla* | 97% | 90% | 87.04% | 4.00E-44 | XR_010132090.1 | |
| | | | | *Macaca fascicularis* | 100% | 90% | 90.07% | 8.00E-46 | CP141342.1 | |
| Table S3A [24] | GACCAAGCAGAAAGAACCGACCAAAAACTGAACAGAACCGACCAAAACTGACCAAAACCGACAAAAAACGACCAAAACCGACCAAGAACCGACTAAGAACCGACTAAGAACTGACCGTATACAGAACCGACCAAAATCAACCAGAACCG | 149 | Ray-finned Fishes : 281 fragments | *Arctogadus glacialis* | 92.00% | 80.00% | 73.60% | **1.00E-19** | OZ177897.1 | N/A |
| | | | | *Tripterygion delaisi* | 92% | 78.10% | 71.85% | 7.00E-16 | OZ285693.1 | |
| | | | | *Zeus faber* | 99% | 77.30% | 76.53% | 9.00E-15 | OY482863.1 | |
| | | | | *Echiichthys vipera* | 92% | 76.39% | 70.28% | 3.00E-14 | OY829646.1 | |

All raw sequencing data have been deposited in PRJNA1309836 and are available for independent re-analysis.

Main observations:

(1) Five maize HSF fragments were identified among green plant sequences. The majority exhibit Ident (α) values distinct from those of modern cultivated maize, excluding recent contamination (maize was absent during the Cretaceous, so the signal must have entered the fossil between formation and excavation). Given that maize was introduced to China from overseas only around the 17$^{th}$ century, this finding suggests that environmental biomolecules may have begun infiltrating and accumulating within the fossil from the 17$^{th}$ century onward. Alternative explanations cannot be ruled out at present; further discrimination requires sedimentological and chronological evidence.

(2) A total of 281 ray-finned fish HSF fragments were identified, including sequences that align with actinopterygian taxa not recorded in the modern regional fauna (marine forms). The regional geological record shows no evidence of marine transgression. Moreover, most fragments display Ident (α) values divergent from modern genomes. A reasonable interpretation is that these derive from extinct species contemporaneous with the fossil deposit (~120 Ma). In an open-system context, this remains a probabilistic inference and alternative sources over geological timescales cannot be excluded. Even so, the finding would have implications for biogeography and phylogenetic reconstruction and therefore merits targeted replication experiments.

(3) Fifty non-human primate HSFs were detected. The corresponding sequences show substantial divergence from the modern human genome, ruling out recent human contamination. These fragments may reflect ancient hominoid molecular input and deposition, providing potential molecular evidence for paleoecology and species turnover, although further validation and replication on additional samples are required.

(4) The HSF workflow combined with preservation-system diagnosis challenges the conventional aDNA framework and highlights a key insight: in open systems, fragment provenance can only be inferred probabilistically and carries inherent uncertainty. As most fossils do not represent closed systems, identifying stable signals amid uncertainty emerges as both the central challenge and the primary opportunity presented by the new methodology.

(5) After excluding artifactual factors, this study identified two previously undocumented classes of large-scale DNA structural patterns that may reflect underlying mechanisms of genome evolution:

> Coding Region Sliding Replication and Recombination (CRSRR): Observed within a transposase-coding region of actinopterygian HSFs, this pattern involves sliding replication coupled with recombination. Given that roughly half of fish genomes consist of transposable elements, with transposase as a core functional component, and that the Cretaceous was a period of rapid ordinal- and family-level radiation in actinopterygians, CRSRR may be associated with large-scale self-modification of the genome and could represent a fundamental driver [24].

> Sequence Reversal and Rearrangement (SRRA): Detected in fragments mapping to regions of ancient *Homo sapiens* genes, this pattern is characterized by the insertion of hairpin-forming sequences within 3'untranslated regions. Composed of multiple insertion – deletion (INDEL) events, such insertions may alter mRNA stability and expression levels without modifying the coding sequence or protein structure, resulting in relatively subtle regulatory effects in modern humans [15].

(6) Deterministic signals: To our knowledge, this study provides the first identification of these two structural classes in paleobiological samples; no equivalent mechanisms have yet been described in extant biology. Regardless of their precise temporal origin (Cretaceous or more recent), these rearrangements have the potential to form novel gene-structural elements capable of influencing protein expression and genome evolution through as-yet-uncharacterized pathways, thereby implicating fundamental regulatory mechanisms of evolutionary change.

(7) These findings may have surprised many researchers, primarily because CRSRR and SRRA fragments do not conform to the classical aDNA criteria — short fragments (≤60 bp) exhibiting terminal deamination and typically preserved for no more than one million years. Under this restrictive framework, similar important discoveries by our group and others have frequently been dismissed as "operational contamination". As a result, they have rarely appeared in high-impact journals and have not been widely recognized by the scientific community. Notably, CRSRR and SRRA are not conventional insertion mutations. Rather, they represent complex mutational events involving small-scale changes (spanning dozens of bases) that combine sliding replication, reverse transcription, tandem duplication, and insertion processes. No corresponding variants have been identified in modern genomes, suggesting that they may represent extinct intermediate forms in evolutionary history. Although the precise underlying mechanisms remain unclear, these structures hold significant biological and theoretical importance.

**Summary**: The above case illustrates that HSF selection can recover molecular signals systematically excluded by conventional pipelines in open-system contexts. The example is intended to demonstrate the methodological boundary and capability of the approach rather than to provide definitive evolutionary conclusions at this stage; its nature is illustrative ("proof of concept") rather than statistically confirmatory.

## Discussion

### Limitations of Conventional Methods and Their Disciplinary Impact

Conventional aDNA methodologies have long treated most fossils, particularly ancient hominin remains, as closed systems containing only endogenous DNA plus modern contamination. This assumption overlooks the reality that the majority function as open containers capable of accumulating molecular mixtures from multiple temporal phases and biological sources over geological timescales. While these methods effectively control modern contamination, they lack systematic mechanisms to exclude ancient exogenous inputs. Even with continuous algorithmic refinement, prevailing pipelines have never incorporated fossil openness or multisource contamination history as explicit analytical prerequisites [25, 26]. Consequently, genome reconstructions from open-system samples (e.g., Denisovan and Neanderthal assemblies) require rigorous evaluation of the potential confounding effects of ancient molecular admixture on sequence purity and inferential reliability.

A particularly concerning practice is the frequent omission of key provenance data in published studies. The critical information, namely the fragment origins underlying assembled genomes (specific read accessions and sequences), is typically buried within massive raw datasets (often gigabyte-scale), complex pipelines, and undocumented parameters. A simple FASTA file of only a few kilobytes would suffice to clearly disclose these origins, yet such files are rarely made publicly available. If provided, these sequences could be verified at essentially no cost through online BLAST searches, enabling rapid and independent validation of published conclusions. Full *de novo* replication, requiring substantial funding, computing resources, and specialized expertise, remains practically infeasible under current funding constraints.

From lower organisms to higher vertebrates, DNA sequence amplification constitutes the foundation of genome expansion and increasing complexity, serving as the primary window for observing molecular evolutionary phenomena. However, conventional approaches, constrained by the concept of "classical aDNA features," typically recover short fragments (≤60 bp). While these are suitable for detecting single-nucleotide polymorphisms (SNPs), they are largely incapable of capturing longer structural variants such as CRSRR and SRRA. As a result, rapid genome expansion and complex rearrangements remain largely invisible, imposing a fundamental methodological limitation on the observable range of evolutionary processes. This constraint not only restricts the development of aDNA techniques and the scope of research questions but also leads to the misclassification of sequences exhibiting amplification or large-scale rearrangement as "operational contamination," thereby impeding the discovery of novel evolutionary mechanisms.

### Conventional Assemblies Represent Multisource Statistical Composites Rather Than Authentic Individual Genomes

In open-system fossils, exogenous DNA can enter continuously: early-infiltrating molecules degrade over time while later inputs accumulate, creating a temporal structure characterized by scarce old DNA and abundant new DNA. The total DNA content may derive from different periods, individuals, or even species, rendering provenance inherently difficult to resolve. Conventional methods cannot logically prove that the recovered DNA originates from the host individual; the resulting assemblies are typically statistical composites of multiple species, individuals, or hominin groups rather than faithful representations of a single genome.

The conventional pipeline is essentially a statistical inference strategy operating under low signal-to-noise conditions. It rests on several critical assumptions: that damage patterns can discriminate sources, that contamination proportions can be accurately estimated, and that multi-sample consistency can validate results. When these assumptions are violated, conclusions become indeterminate.

Key steps include high-coverage sequencing, contamination estimation, and damage validation. However, probe-directed enrichment severely under-represents the true proportion of exogenous contaminant fragments, which often exceed 95% [27]. Consequently, contamination levels are not accurately reflected, and estimates of endogenous coverage are substantially distorted. Selected fragments are predominantly 30~60 bp long, and sequences from modern humans, other archaic hominins, or non-hominin taxa frequently exhibit high similarity in both reference projection and damage profiles, rendering them indistinguishable. Thus, reported "high coverage" is not necessarily authentic; reference bias and parameter choices further introduce mis-mapping risks. Relying on statistical consensus from large numbers of short reads carries inherent methodological vulnerabilities.

Additional lines of evidence, such as multi-sample consistency and population genetic statistics, also rest on fragile premises. Ancient hominin samples are predominantly recovered from Northern Hemisphere caves with similar soil microbiomes, potentially introducing comparable non-host contamination. Prolonged human activity in these regions further elevates the risk of non-host Homo-derived DNA. Consequently, multi-sample SNP concordance does not robustly prove endogenous origin, and population-level statistics lack host specificity.

In recent years, the concept of "classical aDNA features" has given rise to numerous analytical methods. For example, in the metagenomic study of aDNA and ancient pathogens, two representative approaches have been developed within this framework: HOPS and aMeta [28, 29]. We recommend that such methods be applied with caution only after the preservation-system state of the sample has been explicitly characterized. In particular, target-guided alignment should be avoided whenever possible in favor of unrestricted genomic references.

In summary, researchers must remain highly vigilant against logical flaws at the conceptual level: once the foundational assumptions of a methodology are flawed, subsequent experimental steps are unlikely to fully compensate for or correct the resulting bias and may systematically mislead scientific conclusions.

**Value of Conventional Methods**

In open systems, deamination patterns serve as a useful probe of biomolecule-water interactions, although they are insufficient on their own to determine molecular provenance. In closed systems (e.g., permafrost or sealed volcanic ash deposits), they can still yield relatively complete genome reconstructions, provided parasitic and symbiotic contamination is rigorously excluded. However, this comes at several costs: (1) substantial authentic host fragments are frequently discarded; (2) retained fragments are typically short and genetically limited, increasing the risk of misassignment; and (3) the analytical timescale is largely confined to the million-year range, imposing a practical constraint on the exploration of deeper geological time and more ancient targets.

**Scientific Inference Must Rest on Rigorous Logical Premises**

If the foundational premise that "DNA derives from the host individual" cannot be independently verified, no amount of statistical sophistication can compensate for the underlying logical defect. Grounded in first-principles reasoning, this study systematically exposes inherent logical vulnerabilities in conventional methodologies and questions the reliability of certain paleogenomic conclusions drawn over the past two decades. Traditional approaches remain the dominant paradigm, with results frequently published in CNS-level journals and profoundly shaping disciplinary trajectories. In this academic ecosystem, the HSF framework faces substantial barriers to acceptance in high-impact venues and to securing funding.

We emphasize that the primary logical limitation of conventional methods lies in the oversimplified assumption of sample closure. Many published conclusions would likely remain valid if supplemented by independent assessment of preservation-system closure and additional lines of validation. At the same time, open, multisource samples require more conservative analytical approaches. Such caution would enhance the robustness and broader applicability of the overall analytical framework. Should the identified methodological limitations be independently confirmed, a broad range of published paleogenomic inferences, including raw datasets, reconstructed genomes, interpretations of ancient–modern gene flow, and downstream biomedical applications, will require systematic reappraisal and may necessitate partial correction or retraction. This discussion constitutes a systematic methodological critique of the foundational premises of aDNA research, centered on whether DNA provenance can be rigorously verified, rather than an attack on specific experimental results.

**Paradigm Shift in Paleontological Research**

From first principles, we propose a new analytical framework, the HSF Posterior Sourcing Framework. This paradigm treats fossils as dynamic molecular containers and classifies them into four preservation-system types and eight molecular state categories (H/h × D/d × S/s). It shifts the conventional workflow, which relies on statistical inference under restrictive assumptions, into an interdisciplinary posterior framework that integrates paleogeology, physics, statistics, and paleobiology. Inference proceeds along the explicit sequence "physical priors (preservation-system diagnosis) → statistical isolation → phylogenetic consistency," beginning with environmental constraints, followed by statistical independence testing and exclusion, and culminating in phylogenetic consistency validation.

Although still in its early exploratory stage and far from being a mature, ready-to-use solution, the framework has already identified large-scale DNA structural patterns (e.g., CRSRR and SRRA) that are systematically overlooked by conventional pipelines, demonstrating enhanced information recovery capacity and challenging the conventional million-year limit for aDNA recovery. Our intention is to offer a set of conceptual and operational ideas that stimulate constructive criticism, refinement, and possibly the development of entirely different approaches. Rigorous validation will require independent laboratories to test well-characterized closed-system samples and to conduct parallel comparisons with conventional methods in order to objectively assess its effectiveness and relative advantages.

## Outlook

Conventional methods effectively confine aDNA research to timescales within the million-year range. In contrast, the HSF framework offers the potential to overcome this limitation by extending molecular paleontology into deeper geological time through a paradigm-shifting approach. HSF fragments provide distinct operational and representational advantages for species identification, phylogenetic placement, and reconstruction of molecular evolutionary dynamics. Although the availability of recoverable fossils continues to decline, the HSF approach can increase the effective number of research targets by broadening temporal and spatial coverage. We recommend systematic screening of HSFs across ancient strata and diverse substrates, including skeletal fossils, source rocks, archaeological ceramics, permafrost, and fine-grained sedimentary deposits. Lineages should be identified starting from basal taxa and extended outward to infer evolutionary trajectories. Priority should be given to unambiguously closed and well-preserved samples (e.g., permafrost, *in situ* sediments), where high-confidence genome reconstruction or lineage-specific inference can be performed under stringent contamination controls. This strategy aligns in sampling scope with sedimentary aDNA (sedaDNA) research but provides a more comprehensive logical framework. In particular, targeted HSF selection enhances the reliability of phylogenetic inference in environments with substantial exogenous molecular input.

## Materials and Methods

### 1. Relationship Between Fossil Preservation Systems and Eight Molecular Categories

The four preservation systems do not set the probability of presence for the eight molecular categories; they determine each category's relative proportions. HSF selection leverages these proportional imbalances for reverse inference. IVR and 3D X-ray imaging can assess openness versus closedness, while deamination patterns can distinguish closed degradation from closed preservation systems. The relationships can be summarized and inverted by the following conceptual System × Molecule matrix; the expected matrix is:

**Closed-preservation**: Pore connectivity ↓; H/h ↑, h-HSF* ↓; D/d ↓, S/s ↑

**Closed-degradation**: Pore connectivity ↓; H/h ↑, h-HSF* ↓; D/d ↑, S/s ↑

**Open-preservation**: Pore connectivity ↑; H/h ↓, h-HSF** ↑; D/d ↕, S/s ↕

**Invasion-replacement**: Internal voids ↑, pore connectivity ↑; H/h ↓, h-HSF** ↑; D/d ↕; S/s ↕

Notes: h-HSF*: derived from host-associated parasitic/symbiont organisms; h-HSF**: derived from both host-associated organisms and external environmental sources; ↕: uncertain, but coupled with changes in the local molecular environment and time elapsed.

### 2. Concept, Theoretical Basis, and Scope of HSF Selection

**Definition**: HSF (Host-specific fragment) refers to a DNA fragment that, in sequence alignment analysis, matches a particular genus, species, or higher taxonomic unit in a statistically significant and phylogenetically reasonable manner (i.e., its best hit is statistically significantly isolated), and shows a clear gap from homologous sequences of other species. The identification of HSF rests on three pillars:

#### 2.1 Phylogenetic reasonableness

Interspecific phylogenetic relationships have been robustly established by morphology and molecular phylogenetics, providing a biological constraint.

#### 2.1 Statistical isolation

The E-value gap between the top BLAST hit and the next best hit is ≥ 3 orders of magnitude (see step 4); based on sequence alignment algorithms such as BLAST, the match exhibits both significance and uniqueness.

#### 2.3 Representativeness of reference data

Genomes in the reference database should cover all relevant species involved in the evolutionary history of the fossil locality; at minimum, coverage should reach order or family level when genus/species data are insufficient; public databases (e.g. NCBI nr/nt) should be combined to provide the broadest possible cross-species alignment background.

## 3. Scope of Application and Logical Boundaries

HSF selection is applicable to taxa where closely related species have sequenced genomes, phylogenetic relationships are well resolved, clade structure is stable (e.g. most mammals and vertebrates).

For taxa with poor sequencing coverage or unresolved phylogenetic position, the reliability of HSF assignment decreases markedly and must be used with caution; alternatively, HSFs may be screened at higher taxonomic ranks (e.g. family or order level for a marine decapod with limited genus/species data).

## 4. Core Principles

HSF is not simply the “best hit” sequence. It is a match that exhibits statistical isolation under the dual constraints of statistical significance and phylogenetic consistency.

In this study, “statistically significant isolation” is operationally defined through quantitative criteria in Sub1 (confirmatory HSF) and Sub2 (exploratory HSF) steps. The purpose of Sub2 is not to provide a final host assignment, but to avoid losing potential authentic signals in incomplete reference systems.

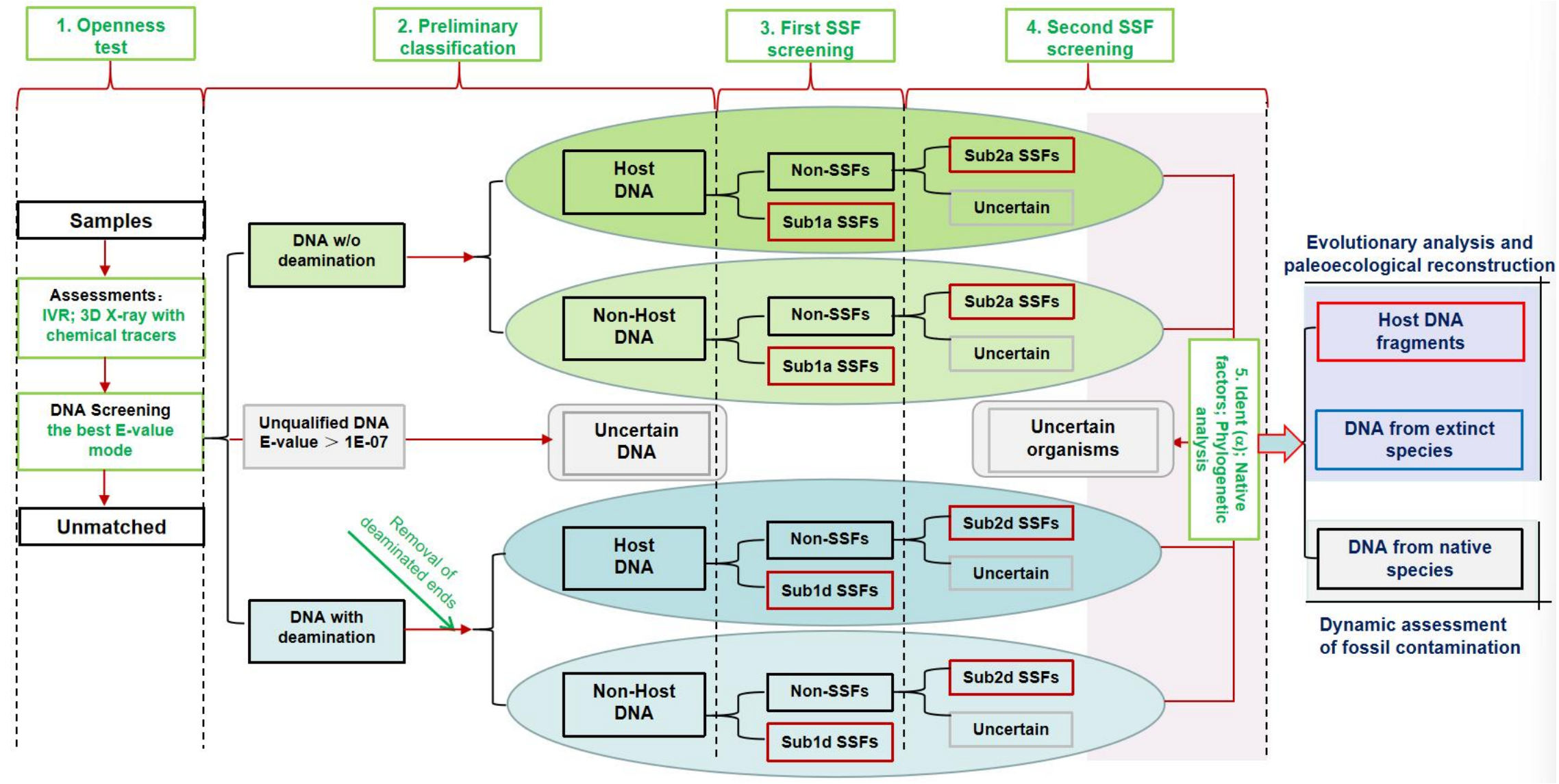

**Figure 3. Workflows for HSF selection**

## 5. HSF Selection Workflows

Figure 3 provides an overview of the entire process: sequence input → alignment filtering → authenticity triage → statistical/phylogenetic testing → uncertainty removal → conditional rescue → output HSF classification.

### 5.1 Reference Databases and E-value Thresholds

Database priority: NCBI nr/nt (to minimize misassignment); E-value threshold: ≤ $1.00 \times 10^{-7}$. This threshold represents an empirical compromise between reducing cross-species misassignment risk and retaining sufficient sensitivity for short fragments in highly complex databases. Constraint: Fragments ≤ 60 bp often fail to reach this threshold and can only serve as auxiliary evidence; they should not be used as core HSF or Sub1 molecules. This step addresses only whether a match is statistically significant, not whether the source is credible.

**5.2 Deamination Status Triage (Branching, Not Classification)**

DNA deamination primarily manifests as C→T transitions enriched at the 3′ end and extending upstream. Therefore, fragments are therefore divided into two classes: d-class (no deamination) → proceed directly to step 3; D-class → trim deamination-enriched terminal bases before re-entering step 1.

Deamination processing must occur before taxonomic assignment, not after. This step restores the original sequence structure and does not constitute an authenticity filter.

**5.3 Functional Group Filtering (Complexity Reduction)**

This step is not performed at the genus or species level, but at broader taxonomic categories (e.g., primates or higher ranks). Based on the preliminary alignment results from step 1, fragments are initially classified into: fragments related to the fossil-bearing taxon group (e.g., primates), fragments unrelated to the fossil-bearing taxon group.

This classification serves only as a complexity-reduction filter for downstream steps 5.4–5.5 and does not constitute evidence for HSF assignment. Only fragments preliminarily associated with the fossil taxon group proceed to statistical and phylogenetic testing.

**5.4 Statistical Criteria for Sub1a/Sub1d HSF (Confirmatory HSF)**

Significant isolation: E-value gap ≥ 3 orders of magnitude between the top hit and the second-best hit (This is an operational empirical threshold that balances sensitivity and specificity; it may be further calibrated in future work.For each fragment, BLAST results are sorted by ascending E-value).

The required statistical gap is: Top 1 E-value vs. Top 2 E-value differs by ≥ 3 orders of magnitude (i.e., Top 2 E-value ≥ 1000 × Top 1 E-value). This criterion excludes cases where the query sequence matches multiple species nearly equally well and ensures that the top hit is statistically isolated rather than merely one of many similar matches.

**5.5 Phylogenetic consistency check (biological constraint)**

The taxonomic assignments of Top 2 and Top 3 hits must be compatible with established phylogenetic relationships. For example: if Top 1 = *Homo sapiens*, then Top 2/Top 3 should belong to closely related taxa (e.g., great apes or other primates); if Top 2/Top 3 belong to distantly related lineages (e.g., bacteria or basal metazoans), phylogenetic consistency is violated.

Only fragments that satisfy both statistical isolation and phylogenetic consistency are classified as Sub1a (undamaged) or Sub1d (deaminated but trimmed) HSF—Type I (high-confidence) assignments. The phylogenetic consistency judgment relies on published consensus phylogenies and is not a per-fragment subjective decision.

**5.6 Types of Assignment Uncertainty — Strict Exclusion**

Fragments that violate phylogenetic expectations are assigned to uncertainty categories (see Table 3) and strictly removed from the Sub1 HSF set:

> **M-type uncertainty (Multiple-hit ambiguity)**: multiple species show essentially identical or near-identical E-values; no unique source can be determined.
>
> **P-type uncertainty (Phylogenetic conflict):** Top 1 matches the expected fossil lineage (e.g., hominin), but Top 2/Top 3 match distantly related organisms (e.g., bacteria). Fast-evolving, undersampled microbial lineages make reliable exclusion difficult; Top 1 cannot be trusted in such cases.

**5.7 AT-type uncertainty (interpretive dilemma at A/T sites)**

A/T positions are ambiguous due to deamination vs. true SNP; such positions should not be used for mutation rate or phylogenetic inference, especially in short, highly deaminated fragments.

**5.8 Low-complexit / repetitive sequences:** excluded outright.

**Table 3. Uncertainty categories**

| Uncertainty Type | Definition | Logical relation to other types | Handling |
|---|---|---|---|
| M-type (Multiple-hit) | E-values essentially equal; no unique source | Can overlap with P-type | Strict exclusion (unless Sub2 rescue) |
| P-type (Phylogenetic conflict) | Top 2/3 distant, violates phylogeny | Can be independent or combined with M | Strict exclusion; check database coverage first |
| AT-type (Interpretive dilemma) | A/T sites unreliable due to deamination | Independent of M/P; often from step 2 | Not used in core assignment; auxiliary only |
| Low-complexity/repeats | — | — | Excluded |

**5.9 Ident (α): a simple quantitative index of fragment–reference similarity**

Definition: Ident (α) = Percent Identity × Query Cover × 100%; example: Identity = 95%, Query Cover = 80% → Ident (α) = 76%.

Ident (α) provides a more comprehensive measure of overall similarity than identity or coverage alone and serves as an auxiliary reference for fragment provenance, but It should not replace phylogenetic judgment. Fragments with Ident (α) ≤ 70% indicate substantial divergence from the reference genome and warrant caution.

**5.10 Sub2 HSF — Conditional Rescue Strategy**

For fragments previously classified as M-type or P-type uncertainty, conditional rescue into Sub2 (exploratory HSF) may be considered if all of the following are satisfied:

The "candidate species" (usually the top 2–3 hits, often lower taxa) responsible for the uncertainty have no documented ecological or geological presence in the fossil region; no Sub1 HSF exists for those "candidate species" in the datasets.

Openness testing (IVR + 3D X-ray) rules out late-stage contamination or mixing.

Ident (α) conform to evolutionary expectations: given a well- represented reference dataset, Ident

(α) reflects matching differences to known genomes and can indicate phylogenetic relationships.

If the sample can be confirmed as a closed system, shows the above data features, and has native factor evidence (e.g., geological, paleoecological, species distribution history), conditional recovery of the fragment is justified.

**5.11 Supplementary Notes**

Preservation system state and fragment assignment are iteratively refined in actual analysis (Bayesian-like feedback loop), not determined unidirectionally.

The top hit is not automatically an HSF.

E-value < 1 is generally biologically meaningful; initial noise filtering uses $\leq 1 \times 10^{-7}$; subsequent isolation requires $\Delta E \geq 3$ orders of magnitude.

In very large databases (hundreds of TB–PB), bit score difference Δbits ≥ 10 (≈ 3 orders of magnitude in E-value) is a widely used practical rule for statistical significance.

Statistical significance ≠ sufficient biological evidence. The NCBI nr/nt database is heavily biased toward model organisms, medically relevant taxa, and certain microbes, while most vertebrate and invertebrate lineages remain undersampled.

For human/primate fragments, Top 1–3 E-values are frequently very close (Δbits < 10) due to over-representation.

When Top 2/3 hits are distant taxa, short conserved domains or database bias must be considered. For undersampled lineages, large Δbits may occur, but precise phylogenetic placement remains difficult → assignment should be conservatively elevated to the appropriate higher taxon.

**5.12 Boundaries of HSF Selection**

HSF directly addresses only one of three independent authenticity dimensions:

**Molecular authenticity** (ancient vs. modern contaminant).

**Lineage authenticity** (most probable phylogenetic source).

**Host authenticity** (whether the molecule derives from the fossil organism itself).

HSF primarily resolves dimension 2 and, when combined with preservation system diagnosis, provides probabilistic support for dimensions 1 and 3 — never deterministic proof.

**HSF + closed system** → high posterior probability that the fragment derives from the host lineage.

**HSF + open system** → high posterior probability that the fragment derives from the surrounding paleoenvironment.

Neither equivalence is absolute proof: HSF + closed systems ≠ deterministic proofs, but rather elevate the subsequent posterior probabilities to a range that can be used for inference.

Comprehensive assessment of three-dimensional molecular authenticity must incorporate native factors (paleogeography, climate evolution, geological events, lineage continuity, ecological dynamics) as boundary conditions and probability constraints.

**5.13 For undersampled clades**

HSF provides evidence of non-random, non-generalized homology matching and membership in a particular phylogenetic branch — not precise species-level assignment. Interpretation is ultimately constrained by reference database completeness.

**5.14 Methodological Summary**

> **Core logical chain:** statistical significance → sequence authenticity triage → phylogenetic consistency → uncertainty removal → conditional rescue. The methodology starts from statistical isolation, applies phylogenetic constraints and rigorous uncertainty filtering, and ultimately yields reliable lineage assignment — consistent with the initial HSF concept.
>
> **Risk mitigation:** Cross-validation across multiple databases (GenBank + custom local databases) offsets reference incompleteness.

All rescue decisions must be accompanied by an independent, traceable evidence chain to minimize subjective bias introduced in Sub2.

## 6. System Characterization of Fossils

**6.1 Openness assessment**

Porosity and volume are quantified using IVR combined with 3D X-ray scanning to evaluate openness and heterogeneity, thereby determining closed versus open preservation state. In the future, this approach may be supplemented by chemical element tracing and taxon-specific sequence tracking techniques to further constrain material exchange history and system state.

**6.2 Signal analysis**

HSF counts and taxonomic distribution are recorded, with h-HSF categories distinguished from host-derived, parasitic/symbiont, and other origins. The distribution of Ident (α) values across HSFs is examined for consistency with expected evolutionary divergence and used to identify lineage signals (interpretive layer).

**6.3 Deamination fragment characteristics**

The number of deaminated fragments is quantified and their taxonomic distribution is tested against expected statistical patterns as evidence of molecular degradation.

**6.4 Background factor evaluation**

Paleogeographic setting, climatic evolution, geological events, lineage continuity, and ecological dynamics are treated as contextual variables. Their influence on biomolecular taxonomic distribution and system characterization is assessed to provide boundary conditions and probabilistic constraints.

**6.5 Important considerations**

Integrated judgment should combine physicochemical characteristics of the embedding matrix,

surface, and sampling region with molecular data. Conclusions based on any single line of evidence should be avoided.

### 6.6 Logical hierarchy

System characterization and HSF selection operate at different levels: system characterization constitutes exogenous constraints, based on depositional environment and input conditions, used to evaluate internal material exchange and information generation mechanisms.

HSF selection is a conditional probabilistic inference performed within these boundaries to parse lineage assignment signals, which cannot substitute for system diagnosis but can serve as supporting evidence (preservation system state is correlated with fragment provenance probability); together, they form a continuous analytical framework from information generation to interpretation.

System diagnosis and provenance inference are iteratively refined rather than unidirectionally determined. Physical determination of the preservation system provides the prior boundary conditions for analysis, while HSF selection constitutes posterior probabilistic inference of lineage origin under those conditions; in causal structure, the relationship is one of “boundary constraint” versus “conditional inference”, not mutual determination.

**Table 4. Diagnostic features of the four preservation systems**

| System | Internal–external exchange fissures | Open porosity/voids | IVR positive | h-HSF⁺ | Deaminated fragments |
|---|---|---|---|---|---|
| **Closed–preservation** | – | – | – | – | – |
| **Closed–degradation** | – | – | – | – | + |
| **Open–preservation** | + | – | + | + | + |
| **Invasion–replacement** | + | + | + | + | + |

**f-HSF⁺ : derived from external environmental species (non-host parasitic/symbiont origin)**

## 7. Software Applications

Sequence alignment is performed using the NCBI Nucleotide BLAST online tool, starting with the Core_nt database and extending to NCBI nr/nt. DNA sequence homology analysis is conducted using MEGA11.

## References

1. Pedersen, M.W. et al. (2015) Ancient and modern environmental DNA. *Philos Trans R Soc Lond B Biol Sci* 370 (1660), 20130383.

2. Arbøll, T.P. et al. (2023) Revealing the secrets of a 2900-year-old clay brick, discovering a time capsule of ancient DNA. *Scientific Reports* 13 (1), 13092.

3. Willerslev, E. et al. (2003) Diverse plant and animal genetic records from holocene and pleistocene sediments. *Science* 300, 791-795.

4. Briggs, A.W. et al. (2007) Patterns of damage in genomic DNA sequences from a Neandertal. *PNAS* 104, 14616-21.

5. Meyer, M. et al. (2012) A high-coverage genome sequence from an archaic Denisovan individual. *Science* 338, 222–226.

6. Burbano, H.A. et al. (2010) Targeted investigation of the Neandertal genome by array-based sequence capture. *Science* 328 (5979), 723-5.

7. Fu, Q. et al. (2025) Denisovan mitochondrial DNA from dental calculus of the >146,000-year-old Harbin cranium. *Cell* 188 (15), 3919-3926 e9.

8. Fu, Q. et al. (2025) The proteome of the late Middle Pleistocene Harbin individual. *Science* 14, 704-707.

9. Kendall, C. et al. (2018) Diagenesis of archaeological bone and tooth. *Palaeogeography Palaeoclimatology Palaeoecology* 491, 21-37.

10. Díaz-Cortés, A. et al. (2024) Diagnosis of archaeological bones: Analyzing the state of conservation of lower Pleistocene bones through diagenesis methods. *Microchemical Journal* 206, 111353.

11. Smith, C.I. et al. (2008) The precision of porosity measurements: Effects of sample pre-treatment on porosity measurements of modern and archaeological bone. *Palaeogeography, Palaeoclimatology, Palaeoecology* 266 (3-4), 175-182.

12. Peterson, J.E. et al. (2010) Influence of microbial biofilms on the preservation of primary soft tissue in fossil and extant archosaurs. *PLoS One* 5 (10), e13334.

13. Saitta, E.T. et al. (2019) Cretaceous dinosaur bone contains recent organic material and provides an environment conducive to microbial communities. *Elife* 8, e46205.

14. Schweitzer, M.H. et al. (2005) Soft-tissue vessels and cellular preservation in Tyrannosaurus rex. *Science* 307, 1952-1955.

15. Wiemann, J. et al. (2018) Fossilization transforms vertebrate hard tissue proteins into N-heterocyclic polymers. *Nat Commun* 9 (1), 4741.

16. Massilani, D. et al. (2022) Microstratigraphic preservation of ancient faunal and hominin DNA in Pleistocene cave sediments. *Proc Natl Acad Sci U S A* 119 (1), e2113666118.

17. Zhao, L.J. et al. (2025) Ancient DNA Reveals Hominoid Evolution: Intermediate DNA Sequences and Advances in Molecular Paleontology. *bioRxiv* 04.27.650833

18. An, R. et al. (2014) Non-enzymatic depurination of nucleic acids: factors and mechanisms. *PLoS One* 9 (12), e115950.

19. Frederico, L.A. et al. (1990) A sensitive genetic assay for the detection of cytosine deamination: determination of rate constants and the activation energy. *Biochemistry* 29 (10), 2532-7.

20. Loeb, L.A. and Preston, B.D. (1986) Mutagenesis by apurinic/apyrimidinic sites. *Annu Rev*

*Genet* 20, 201-30.
21. Krause, J. et al. (2010) The complete mitochondrial DNA genome of an unknown hominin from southern Siberia. *Nature* 464:894-7.
22. Prüfer, K. et al. (2014) The complete genome sequence of a Neanderthal from the Altai Mountains. *Nature*. 505:43-9.
23. Zhang, D. et al., (2020) Denisovan DNA in Late Pleistocene sediments from Baishiya Karst Cave on the Tibetan Plateau. *Science* 370, 584-587.
24. Zhao, W. Q. et al. (2024) Ancient DNA from 120-Million-Year-Old Lycoptera Fossils Reveals Evolutionary Insights. *arXiv* 2412.06521
25. Al-Asadi, H. et al. (2019) Inference and visualization of DNA damage patterns using a grade of membership model. *Bioinformatics*. 35:1292-1298.
26. Ravishankar, S. et al. (2025) Filtering out the noise: metagenomic classifiers optimize ancient DNA mapping. *Briefings in Bioinformatics* 26, bbae646.
27. Fu, Q. et al. (2013) DNA analysis of an early modern human from Tianyuan Cave, China. *Proc. Natl. Acad. Sci. U.S.A*. 110 (6), 2223-2227.
27. Hübler, R. et al. (2019) HOPS: automated detection and authentication of pathogen DNA in archaeological remains. *Genome Biol* 20, 280.
28. Pochon, Z. et al. (2023) aMeta: an accurate and memory- efficient ancient metagenomic profiling workflow. *Genome Biology* 24:242.